# Spectral remapping of natural signals

Md. Shoaibur Rahman
Baylor College of Medicine
Houston, TX 77054
shoaibur.rahman@bcm.edu

**Abstract**

*Here we present an algorithm to procedurally remap spectral contents of natural signals. The algorithm takes in two inputs: a signal whose spectral component needs to be remapped and a warping or remapping function. The algorithm generates one output, which is a remapped version of the original signal. The input signal is remapped into the output signal in two steps. In the analysis step, the algorithm performs a series of operations to modify the spectral content, i.e., compute the warped phase of the signal according to the given remapping function. In the synthesis step, the modified spectral content is combined with the envelope information of the input signal to reconstruct the warped or remapped output signal.*

1. Introduction

This paper presents an algorithm to procedurally remap spectral contents of natural signals. Procedural remapping of natural signals' spectra is of interest in numerous applications. For example, computer music composers are interested in pitch transposition by modifying spectra of a wide variety of sound signals. Systematic pitch transpositions could produce many of our desired perceptual effects of sound signals. To achieve these perceptual effects, efficient remapping of spectral contents of sound signals has been a great importance in audio applications for a long time (Grey and Moorer, 1977; Grey and Gordon, 1978; Samson et al., 1997; Krumhansl and Iverson, 1992; Alluri and Toiviainen, 2010; Vos and Rasch, 1881; Deutsch, 2013). Analogous applications of spectral remapping have been reported in other fields as well, e.g., in clinical applications using EEG (Chemin et al. 2018; Wang et al. 2016) and fMRI (Meszlényi et al. 2017), acoustic oceanographic measurements (Niu et al. 2014; Bonnel et al. 2017), and speech recognition (Chang et al., 2017; Goupell et al., 2008; Noda, 1988; Lee and Rose, 1998).

The process of spectral remapping is equivalent to the warping of each frequency in the signal. So, spectral remapping can be achieved by warping each frequency of the signal based on a given remapping function (**Fig. 1**). The idea of frequency warping was initially proposed in terms of phase vocoders (Flanagan and Golden, 1966; Moorer 1978). Phase vocoder was originally developed for timescale modification of audio and was implemented using filter-banks and Fourier transforms (Dolson, 1986). Then other versions of phase vocoders were reported (Puckette, 1995; Laroche and Dolson, 1999). Although the implementation techniques of the later versions were different from the initial ones, the goal was the same – to change the temporal characteristics of a sound by stretching or compressing the time-base of the spectrogram. These methods introduce errors in the spectral remapping process (Puckette and Brown, 1998). Further efforts had been made to develop more general methods for frequency warping, for instance, using arbitrary allpass maps (von Schroeter, 1999), and more recently, using short time Fourier transforms and Gabor frames (Evangelista and Cavaliere, 2007; Mejstrik and Evangelista, 2016; Wabnik et al., 2005). These methods perform well for remapping spectral contents of simple signals like pure tones or even frequency sweeps. However, these algorithms are susceptible to high spectral leakage during the remapping process for complex natural signals like human speech.

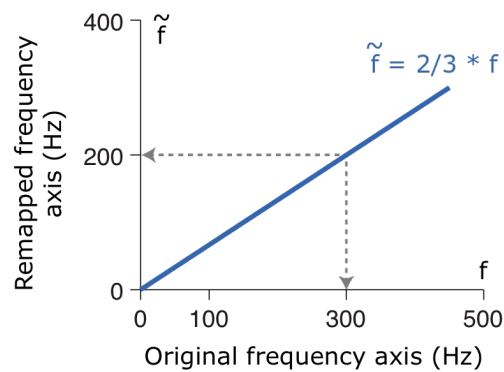

Figure 1. Frequency remapping explained with a warping or remapping function, $\tilde{f} = 2/3 * f$, where f and $\tilde{f}$ are the frequencies of the original and remapped signals, respectively. So, a frequency of 300Hz in original axis will be projected to 200Hz in remapped axis, and vice versa.

To address this gap, here we present an algorithm of frequency warping to remap spectral contents of various signals, including pure tones, frequency sweeps, and a wide variety of natural signals. We have shown that although the algorithm concedes a nominal spectral leakage, it can faithfully remap the spectra of the signals to the desired ones. The validity of the remapped signals was tested using signals of different complexity levels, i.e., from pure tones to human voice with instruments.

2. Algorithm

A high-level description of the proposed algorithm is presented in the block diagram (**Fig. 2**). The algorithm involves analysis and synthesis steps. The analysis step computes the envelope and the warped phases of the input signal. The original phase information is warped using the given remapping or warping function. The synthesis step combines the envelope and the warped phases to reconstruct the output signals. A summary of the algorithm is shown in **Table 1**. The details of the inputs, analysis/synthesis steps, and the output are explained in the following subsections.

2.1. Inputs

The algorithm requires two inputs: first, a real signal $x[n]$ in the time domain with a length of $M \in \mathbb{Z}_+$ with n being the time index; second, a frequency warping function, $\tilde{f}[n] = \text{func}(f[n])$, where $f[n]$ and $\tilde{f}[n]$ are the instantaneous frequencies of the input and output signals, respectively.

2.2. Analysis

*2.2.1. Mean signal, operation signal and analytic signal*: Analysis step starts with computing mean $\mu$ (**Eqn. 1**) of the input signal. The mean signal is subtracted from the original input signal to compute operation signal $s[n]$ (**Eqn. 2**). Hilbert transform (HT) is applied on the operation signal (**Eqn. 3**).

$$\mu = \frac{1}{M} \sum_{k=1}^{M} x[k] \quad (1)$$

$$s[n] = x[n] - \mu \quad (2)$$

$$\hat{s}[n] = \mathcal{H}(s[n]) \quad (3)$$

where, $\mathcal{H}$ represents an operator for HT.

The operation signal $s[n]$ and HT of $\hat{s}[n]$ are used to form the analytical signal $s_a[n]$ (**Eqn. 4**), which is a complex signal in time domain and is defined as:

$$s_a[n] = s[n] + j\hat{s}[n] \quad (4)$$

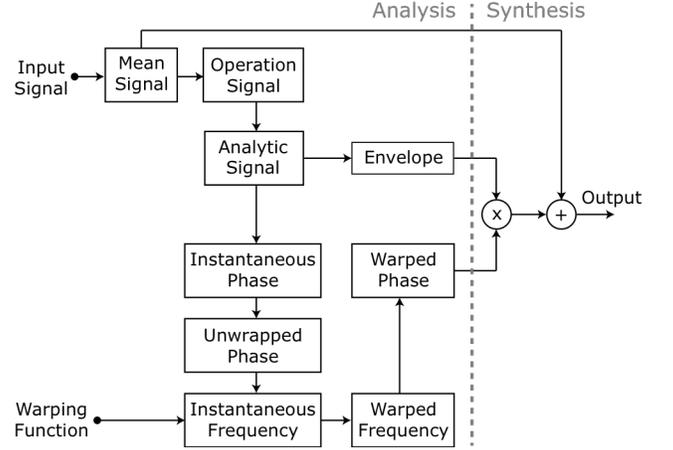

Figure 2. Block diagram representation of the frequency warping algorithm. The input signal is used to compute the mean signal and the operation signal. The analytic signal is computed using the Hilbert transform on the operation signal. The analytic signal provides information about the envelope and instantaneous phase of the signal. The instantaneous phase is unwrapped to obtain instantaneous frequency. A warping function is applied to transform instantaneous frequency into warped frequency, which is converted into warped phase. The envelope and the warped phases are multiplied (element-wise product), which is then added with the mean signal to obtain the warped output signal. The vertical dashed line indicates the separation between the operations during analysis and synthesis steps.

*2.2.2. Envelope and instantaneous phase*: The analytic signal is used to extract envelope information $e[n]$ (**Eqn. 5**) and instantaneous phase $\phi[n]$ (**Eqn. 6**).

$$e[n] = \sqrt{|s[n]|^2 + |\hat{s}[n]|^2} \quad (5)$$

$$\phi[n] = \arctan\left(\frac{\hat{s}[n]}{s[n]}\right) \quad (6)$$

Phase information is represented in radians. In the subsequent operations in the analysis step, only phase information is used. The envelope information remains intact, and is used during the synthesis step.

*2.2.3. Unwrapped phase, instantaneous frequency, warped frequency and warped phase*: The instantaneous phases are converted into warped phases by applying a series of unwrap/wrap and differential/integral operations (**Eqn. 7-10**) (Shoaib et al. 2010; Rahman 2011; Rahman and Haque 2012). The phases can be wrapped between $-\pi$ and $\pi$ for every signal. For instance, the computed phases of $\sin(2\pi + \pi/3)$ and $\sin(-4\pi + \pi/4)$ are $\pi/3$ and $\pi/4$ instead of their actual phases $2\pi + \pi/3$ and $-4\pi + \pi/4$, respectively. So, before any operation being applied, the actual phases are restored by adding appropriate multiples

Table 1. Summary of the algorithm

**Algorithm:** Frequency warping
**Input:** Real signal $\mathbf{x}[n]$ with a signal length of $M \in \mathbb{Z}_+$
**Output:** Estimates of frequency warped signals
**Analysis:**
  1: Construct analytic signal
     1a: Compute mean signal, $\mu$
     1b: Compute operation signal, $s[n] = x[n] - \mu$
     1c: Compute Hilbert transform, $\hat{s}[n]$
     1d: Form analytic signal, $s_a[n] = s[n] + j\hat{s}[n]$
  2: Extract properties of the analytic signal
     2a: Extract envelope, $e[n] = |s_a[n]|$
     2b: Extract instantaneous phase, $\phi[n] = \angle s_a[n]$
  3: Warp frequency and phase
     3a: Unwrap phase, $\theta[n] = \mathcal{U}(\phi[n])$
     3b: Obtain instantaneous frequency, $f[n]$
     3c: Apply frequency warping function and obtain warped frequency, $\tilde{f}[n]$
     3d: Convert $\tilde{f}[n]$ into warped phase, $\tilde{\phi}[n]$
**Synthesis:**
  4: Reconstruct warped output signal
     4a: Perform dot product between the envelope and the cosine of wrapped phases
     4b: Add the product to the mean signal and obtain the warped output signal, $\tilde{x}[n]$

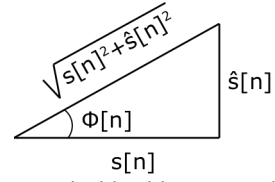

Figure 3. Trigonometric identities. To make this triangular representation analogous to equation (6), base and height must be $s[n]$ and $\hat{s}[n]$, respectively. Accordingly, the hypotenuse is $\sqrt{s[n]^2 + \hat{s}[n]^2}$, which is equal to the envelope $e[n]$. Thus, $\arctan\left(\frac{\hat{s}[n]}{s[n]}\right) = \phi[n] = \arccos\left(\frac{s[n]}{e[n]}\right)$.

of $2\pi$ to the instantaneous phases computed in equation (6). This process is called phase unwrapping. Let's denote the unwrapped phases as $\theta[n]$ and $\mathcal{U}$ be the unwrap operator, then the instantaneous phases are unwrapped as (**Eqn. 7**):

$$\theta[n] = \mathcal{U}(\phi[n]) \qquad (7)$$

$\theta[n]$ are then converted into instantaneous frequencies $f[n]$ by applying a differential operation (**Eqn. 8**), i.e.,

$$f[n] = \frac{F_s}{2\pi}(\theta[n+1] - \theta[n]) \qquad (8)$$

where, $F_s$ denotes sampling frequency. The instantaneous frequencies are modified or warped based on the frequency warping function $\tilde{f}[n] = \text{func}(f[n])$. The warped frequencies, $\tilde{f}[n]$, are converted back to warped phases $\tilde{\theta}[n]$ by applying an integral operation (**Eqn. 9**), i.e.,

$$\tilde{\theta}[n] = \frac{2\pi}{F_s} \sum_{k=1}^{n} \tilde{f}[k], \qquad n = 1, 2 \ldots M \qquad (9)$$

where, $\tilde{\theta}[n]$ are the warped phases. Now, before the warped phases can be used to reconstruct the warped signal at the synthesis step, they must be wrapped between $-\pi$ and $\pi$ by subtracting appropriate multiples of $2\pi$ from $\tilde{\theta}[n]$. Let's denote the wrapped phases as $\tilde{\phi}[n]$, which is computed as (**Eqn. 10**):

$$\tilde{\phi}[n] = \mathcal{U}(\tilde{\theta}[n]) - 2\pi \left\lfloor \frac{\mathcal{U}(\tilde{\theta}[n]) + \pi}{2\pi} \right\rfloor \qquad (10)$$

where, $\lfloor . \rfloor$ denotes a floor operation.

2.3. Synthesis

In the synthesis step, the warped output signal is reconstructed. To execute this step, an element-wise dot product between the envelope and the cosine of wrapped phases is computed. The results of this product are added up with the mean signal to reconstruct the frequency-warped output signals, $\tilde{x}[n]$, (**Eqn. 11**) i.e.,

$$\tilde{x}[n] = \mu + e[n].\cos \tilde{\phi}[n] \qquad (11)$$

To understand how equation (11) works at the synthesis step, here we give an example of perfect reconstruction, i.e., we set up an example so the warped output signal will be identical to the original input signal. An identity frequency warping function, i.e., $\tilde{f}[n] = f[n]$, is an obvious choice for the perfect reconstruction. With the identity frequency warping function, we find that the wrapped phases for original input signal and warped output signal are identical, i.e., $\tilde{\phi}[n] = \phi[n]$, after applying unwrap/wrap and differential/integral operations in the analysis steps. So, according to the equation (11), the signal can be reconstructed as:

$$\tilde{x}[n] = \mu + e[n].\cos \phi[n] \qquad (12)$$

Using trigonometric identities (**Fig. 3**):

$$\begin{aligned}\tilde{x}[n] &= \mu + e[n].\cos\arctan\left(\frac{\hat{s}[n]}{s[n]}\right) \\ &= \mu + e[n].\cos\arccos\left(\frac{s[n]}{e[n]}\right) \\ &= \mu + e[n].\frac{x[n] - \mu}{e[n]} \\ &= x[n]\end{aligned} \qquad (13)$$

Therefore, with an identity frequency warping function, the warped output signal is identical to the original input signal. This verifies that equation (11) can be used to estimate the warped output signal.

### 2.4. Output

The algorithm returns a single output, which is the frequency-warped signal of the original input signal. Since the envelope of the input signal remained intact in the analysis steps and is just used in the synthesis step, the output signal maintains almost the same envelope properties that were in the original input signal.

### 3. Test and validation of the algorithm

To test and validate the performance of the algorithm, we considered examples with three types of signals: pure tone sinusoid, frequency sweep sinusoid, and natural sound signals. For simplicity and to keep the analyses consistent across the examples, the same warping function was used in all examples. Specifically, we used the following simple linear function for frequency warping:

$$\tilde{f}[n] = 2/3 * f[n] \qquad (14)$$

This implies the following inverse frequency warping or de-warping function:

$$f[n] = 3/2 * \tilde{f}[n] \qquad (15)$$

### 3.1. Validation with pure tone sinusoid

A pure tone sinusoidal input signal was generated with a sampling frequency of 16kHz. The frequency of the signal was 300Hz (**Fig. 4A**). The power spectrum (PS) of the signal confirmed that its frequency was 300Hz, i.e., the power of the signal was concentrated around 300Hz (**Fig. 4B**). The signal and the warping function were used as the inputs to the frequency-warping algorithm, which outputted a sinusoidal signal of a different frequency (**Fig. 4C**). Theoretically, according to the equation (14), the frequency of the warped output signal should be: $\tilde{f} = 2/3 * 300 = 200$Hz. PS analysis confirmed that the frequency of the warped signal was 200Hz (**Fig. 4D**). Now, for a lossless algorithm, if the warped signal and the de-warping function are used as the inputs to the algorithm, then we should obtain an output that is identical to the original input signal. Hence, according to equation (15), the frequency of the de-warped signal must be: $f = 3/2 * 200 = 300$Hz. The de-warped signal was estimated (**Fig. 4E**). The cross-correlation between this estimated de-warped and the original signals was unity, i.e., $r = 1$. Additionally, PS analysis confirmed that the frequency of the de-warped signal was 300Hz, i.e., the power spectra of the de-warped and original signal were highly correlated ($r = 0.978$) (**Fig. 4F**). A correlation value of less than 1 implies that the algorithm concedes some spectral leakage during the warping. However, because of the high correlation between the original and de-warped signals (and between their power spectra), the leakage can be negligible.

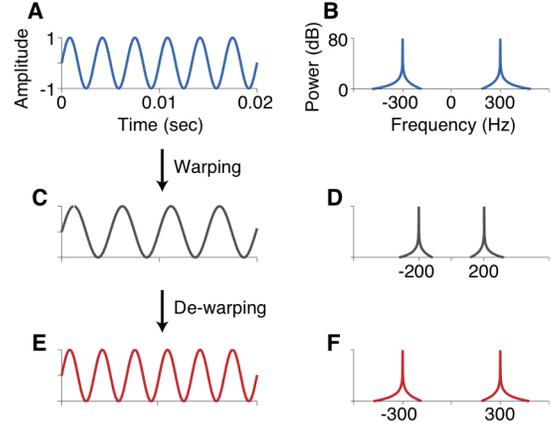

Figure 4. Validation of frequency warping algorithm with a pure tone sinusoid. **A**) Pure tone sinusoidal input signal with a frequency of 300 Hz. Original signal was 1 second long; a shorter version is shown for better visualization. **B**) Power spectrum density (PSD) of the input signal. **C**) Warped signal obtained after applying frequency warping function on the input signal. **D**) PSD of warped signal. **E**) De-warped signal obtained after applying inverse warping function on warped signal. **F**) PSD of de-warped signal. For a lossless algorithm, de-warped signal presented in E must be same to the original input presented in A, therefore, PSD presented in F must be same to that in B.

### 3.2. Validation with frequency sweep sinusoid

A frequency sweep sinusoidal input signal was generated with a sampling frequency of 16kHz. The frequency o the signal changed logarithmically from 21Hz to 480Hz in 1 second (**Fig. 5A**). PS analysis confirmed that the start and the end frequencies of the signal were 21Hz and 480Hz, respectively (**Fig. 5B**). The signal and the warping function were used as the inputs to the algorithm, which outputted a warped frequency sweep sinusoidal signal, whose start and end frequencies were different from that in the original signal (**Fig. 5C**). Theoretically, according to equation (14), the start and end frequencies should be 14Hz and 320Hz. PS analysis confirmed that the start and end frequencies of the warped signal were 14Hz and 320Hz (**Fig. 5D**). Now, to check for the spectral leakage, the warped signal and the de-warping function were used as the inputs to the algorithm, which outputted the de-warped signal (**Fig. 5E**), which was similar to the original signal ($r = 0.998$). Also, their power spectra were alike (r = 0.996); hence, the start and the end frequencies of this de-warped signal were 21Hz and 480Hz, respectively (**Fig. 5F**).

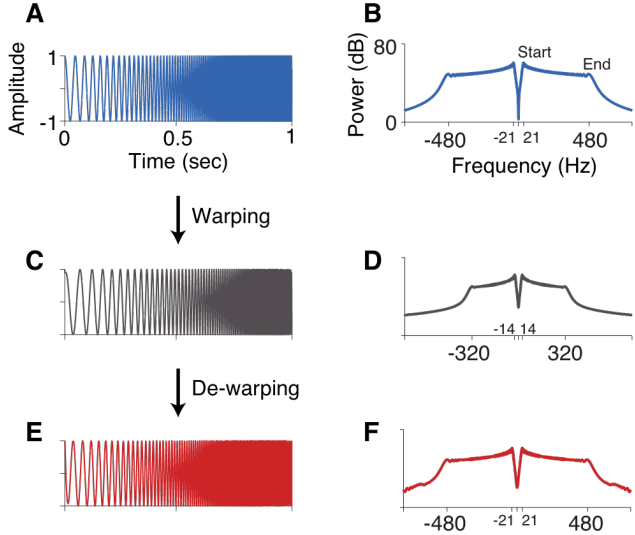

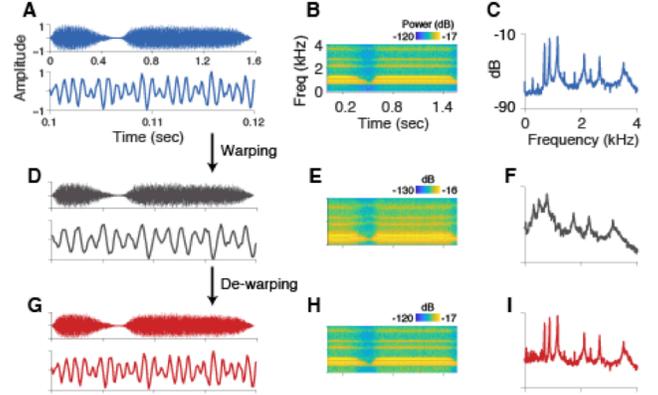

Figure 5. Validation of frequency warping algorithm with a frequency sweep sinusoid. **A**) Frequency sweep sinusoidal input signal, whose frequency increases logarithmically from 21Hz to 480Hz in 1 second. **B**) PSD of the input signal showing the start and end frequencies. **C**) Warped signal obtained after applying frequency warping function on the input signal. **D**) PSD of warped signal. **E**) De-warped signal obtained after applying inverse warping function on warped signal. **F**) PSD of de-warped signal. For a lossless algorithm, de-warped signal presented in E must be same to the original input signal presented in A, therefore, PSD presented in F must be same to that in B.

Figure 6. Validation of the algorithm with more common natural signal. **A**) Top: Original input signal of train whistles (train.mat, Fs = 8192Hz, 1.57 seconds). Bottom: Input signal zoomed-in between 0.1 and 0.12 seconds. **B**) Time-frequency plot of power distribution of input signal. **C**) PSD of the input signal. **D**) Warped signal (top: entire duration, bottom: zoomed-in version). **E**) Time-frequency plot of power distribution of warped signal. **F**) PSD of the warped signal. **G**) De-warped signal (top: entire duration, bottom: zoomed-in version). **H**) Time-frequency plot of power distribution of de-warped signal. **I**) PSD of the de-warped signal.

### 3.3. Validation with natural signals

Seven natural signals were used to validate the algorithm. The signals were available in Matlab2018b either in .mat or .wav formats. A description of the signals is listed in **Table 2**.

A train whistles signal was used as the input to the algorithm (**Fig. 6A**). The signal comprised several major frequency bands as displayed in power distribution over time and frequency (**Fig. 6B**). Alternatively, the power of the signal was concentrated at certain frequencies as indicated by the larger peaks on the power spectrum of the signal (**Fig. 6C**). The signal and the warping function were used as the inputs to the algorithm, which outputted a warped signal. The overall envelope of the warped signal remained intact but the frequencies were changed as expected (**Fig. 6D**). The power of the warped signal essentially shifted to lower frequency bands as shown in its power distribution over time and frequency (**Fig. 6E**). Indeed, the frequencies of the original signal were reduced by a factor of 3/2, which is more obvious on the frequency spectrum (**Fig. 6F**). Now, to check for the spectral leakage, the warped signal and the de-warping function were used as the inputs to the algorithm, which outputted the de-warped signal (**Fig. 6G**). We found that the de-warped signal was nearly identical to the original input signal (r = 0.99). Moreover, the distribution of power over time and frequency was shifted back to the higher frequency bands (**Fig. 6H**). More precisely, the frequencies of the warped signal were increased by a factor of 3/2, which was more obvious on the power spectrum of the de-warped signal (**Fig. 6I**). This shift in frequency made the spectrum of the de-warped signal similar to that of the original signal (r = 0.851).

Similar analyses were performed with the six remaining example signals. We found that frequency spectra of the original input signal and the de-warped signals were highly correlated (**Fig. 7A-F**). High correlation values in all these examples demonstrate that although the algorithm concedes some spectral leakage, it performs well for a frequency warping for a wide variety of natural signals, including vocals with instruments.

Table 2. Natural signals used for validation

| Signals (name and format available in Matlab2018b) | Fs (Hz) | Duration (sec) |
|---|---|---|
| 1. Whistles of a train (train.mat) | 8192 | 1.57 |
| 2. Chirping of a bird (chirp.mat) | 8192 | 1.5 |
| 3. Hallelujah chorus (handel.mat) | 8192 | 8.92 |
| 4. Fall and splat sound (splat.mat) | 8192 | 1.22 |
| 5. Sound of laughing (laughter.mat) | 8192 | 6.42 |
| 6. Male voice speaking (speech_dft_8kHz.wav) | 8000 | 4.99 |
| 7. Rock with vocals, drums and guitar (audio48kHz.wav) | 48000 | 8.96 |

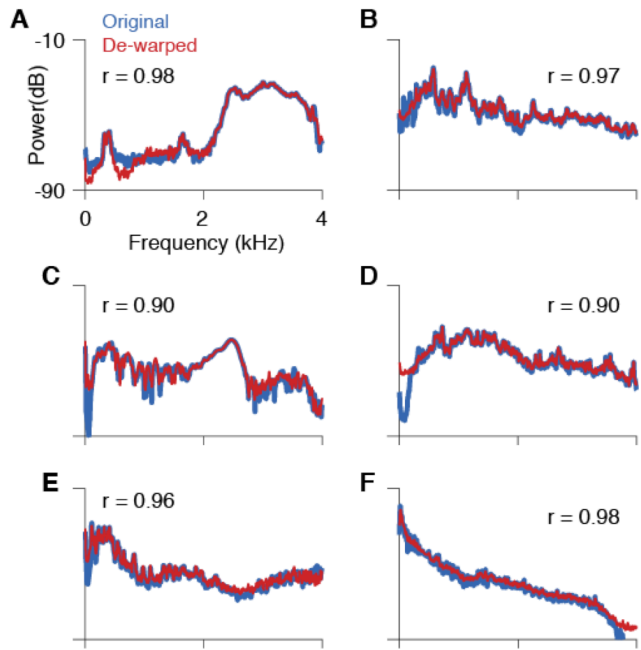

Figure 7. Validation with different natural signals. Power spectrum of: **A**) Chirping of a bird. **B**) Hallelujah chorus from Handel's Messiah. **C**) Fall and splat sound. **D**) Sound of laughing of humans. **E**) Male voice speaking. **F**) Rock with vocals, drums and guitar (used Fs = 8000, 53.78 seconds to keep consistent with sampling frequencies of the other signals in this figure. Conclusions are not changed when using Fs = 48000Hz).

4. Discussion

The algorithm can faithfully warp the frequencies of diverse signals with various levels of complexities, i.e., from pure tones to human voice with instruments. However, it also subjects to a number of limitations. First, the algorithm relies on the estimation of instantaneous frequencies. So, without a good estimation of the instantaneous frequency, the algorithm will perform poorly. The method of estimating frequencies used in this paper has widely been studied, but a better method will increase the performance of the algorithm. Second, spectral leakage is higher for complex signals as compared to simple signals like pure tones. Again, the estimation of instantaneous frequencies of the complex signals is less accurate. This error in estimation may be accounted for as the spectral leakage. However, the spectral leakage was nominal. Third, the computational complexity of the algorithm has not been assessed in this paper. The algorithm performs pretty fast in general as it uses FFT to compute Hilbert transform in the analysis step. For instance, each example in this paper took a fraction of seconds to perform warping and de-warping using Matlab 2018b in a MacOSX (1.4 GHz, Intel Core i5, 4GB RAM). Larger signals may be partitioned in time and each partition can be used as a separate input to make the computation faster.

This algorithm can be applied to transform perceptual signals from one modality to the other, e.g., from audio frequency to tactile and vice versa (Rahman et al. 2017; Convento et al. 2018; Rahman and Yau 2019; Rahman et al. 2019). These applications will potentially provide us new insights to understand perceptual links between different sensory systems.